\documentclass[twocolumn,aps,prd,superscriptaddress,nofootinbib]{revtex4-2}

\usepackage{amsmath,amssymb,graphicx}
\usepackage{hyperref}
\usepackage{mathptmx}
\usepackage{booktabs}
\usepackage{comment}

\begin{document}

\title{Secure Medical Data Transmission Using Quantum Key Distribution and Post-Quantum Cryptography in Real-World Fiber Networks}

\author{Vasile-Laurențiu~Dosan}
\email{l.dosan@qo-jena.com}
\thanks{This author contributed equally to this work.}
\affiliation{Quantum Optics Jena GmbH, Am Zementwerk~8, 07745 Jena, Germany}
\affiliation{Institute of Applied Physics, Friedrich Schiller University Jena, Albert-Einstein-Str.~6, 07745 Jena, Germany}
\affiliation{Max Planck School of Photonics, Friedrich Schiller University Jena, Albert-Einstein-Str.~15, 07745 Jena, Germany}

\author{Paul~Spooren}
\email{paul.spooren@hs-nordhausen.de}
\thanks{This author contributed equally to this work.}
\affiliation{University of Applied Sciences Nordhausen, Weinberghof 4, 99734 Nordhausen, Germany}


\author{Sebastian~Moeckel}
\affiliation{University of Applied Sciences Nordhausen, Weinberghof 4, 99734 Nordhausen, Germany}


\author{Alessandro~Zannotti}
\affiliation{Quantum Optics Jena GmbH, Am Zementwerk~8, 07745 Jena, Germany}


\author{Alek~Lagarrigue}
\affiliation{Quantum Optics Jena GmbH, Am Zementwerk~8, 07745 Jena, Germany}

\author{Pablo~Vazquez}
\affiliation{Quantum Optics Jena GmbH, Am Zementwerk~8, 07745 Jena, Germany}

\author{Marc~Bodenstein}
\affiliation{University of Applied Sciences Nordhausen, Weinberghof 4, 99734 Nordhausen, Germany}

\author{Jonas~Jelonek}
\affiliation{University of Applied Sciences Nordhausen, Weinberghof 4, 99734 Nordhausen, Germany}


\author{Sarika~Mishra}
\affiliation{Fraunhofer Institute for Applied Optics and Precision Engineering IOF, Albert-Einstein-Straße~7, 07745 Jena, Germany}

\author{Jansen~Dwan}
\affiliation{Fraunhofer Institute for Applied Optics and Precision Engineering IOF, Albert-Einstein-Straße~7, 07745 Jena, Germany}


\author{Natasa~Pavlovic~Tucakovic}
\affiliation{Fraunhofer Institute for Applied Optics and Precision Engineering IOF, Albert-Einstein-Straße~7, 07745 Jena, Germany}

\author{Fabian~Steinlechner}
\affiliation{Institute of Applied Physics, Friedrich Schiller University Jena, Albert-Einstein-Str.~6, 07745 Jena, Germany}
\affiliation{Max Planck School of Photonics, Friedrich Schiller University Jena, Albert-Einstein-Str.~15, 07745 Jena, Germany}
\affiliation{Fraunhofer Institute for Applied Optics and Precision Engineering IOF, Albert-Einstein-Straße~7, 07745 Jena, Germany}

\author{Kevin~Füchsel}
\affiliation{Quantum Optics Jena GmbH, Am Zementwerk~8, 07745 Jena, Germany}

\author{Thomas~Hühn}
\affiliation{University of Applied Sciences Nordhausen, Weinberghof 4, 99734 Nordhausen, Germany}

\author{Oliver~de~Vries}
\affiliation{Quantum Optics Jena GmbH, Am Zementwerk~8, 07745 Jena, Germany}

\begin{abstract}

The threat quantum computers pose to classical public-key cryptography motivates the deployment of quantum-safe communication for critical infrastructure such as healthcare, finance, and energy systems. Quantum key distribution (QKD) and post-quantum cryptography (PQC) offer complementary security guarantees, information-theoretic key exchange and quantum-resistant end-to-end authentication that can be combined in a layered architecture. Here, we demonstrate a field-deployed quantum-secure network integrating entanglement-based QKD with end-to-end PQC over 140 km of installed fiber in Thuringia, Germany, connecting a rural health kiosk to a university hospital via a trusted-node architecture comprising heterogeneous underground and aerial fiber links. Unlike conventional deployments that rely on a dedicated key management system to forward keys to applications, our architecture injects QKD keys directly into standard Linux-based virtual private network (VPN) tunnels between adjacent nodes, while PQC secures the communication end-to-end, remaining fully compatible with existing infrastructure and software. Polarization-entangled photon pairs were generated at 810 nm and 1550 nm, with the telecom photon transmitted over deployed fiber. Active polarization stabilization and dispersion compensation preserve the entanglement and enable 22 days of continuous, fully autonomous operation, further underscoring the technological maturity of entanglement-based QKD approaches in a real-world fiber environment. Although the two deployed links were operated during separate rather than concurrent periods, the predominantly aerial link exhibited markedly greater instability, with QBER variations most strongly correlated with wind speed. The generated keys secured a telemedicine proof-of-concept without modifying existing medical systems, demonstrating a practical framework for quantum-safe critical infrastructures.

\end{abstract}

\maketitle

\section{Introduction}

The anticipated capabilities of quantum computers threaten the security of classical public-key cryptography, motivating the development of quantum-safe communication technologies \cite{mosca2018cybersecurity}. Quantum key distribution (QKD) enables information-theoretically secure key exchange based on the principles of quantum mechanics \cite{bennett1984quantum, ekert1991quantum, gisin2002quantum}. In contrast, post-quantum cryptography (PQC) is an algorithm-based approach that derives its security from computational problems that are believed to be intractable even for quantum computers \cite{nistpqc}. Together, these approaches are driving real-world deployments of secure communication networks \cite{peev2009secoqc, chen2021integrated}, particularly for applications requiring long-term data confidentiality.

QKD enables a range of concrete application-level use cases that have recently been demonstrated in real-world scenarios. A prominent example is quantum-secured digital payments, where quantum states are used to generate inherently unforgeable cryptographic tokens (“quantum cryptograms”) that guarantee transaction integrity and user privacy even in untrusted networks \cite{quantum_payment_2023}. In addition, QKD can be applied to secure authentication and message integrity, where quantum-generated keys protect authentication tags or signatures against forgery, even by adversaries with unlimited computational power \cite{lucamarini2018overcoming}.

At the network level, QKD is increasingly being deployed for end-to-end encryption of sensitive data streams, such as inter-data center communication or backbone links, where it provides long-term confidentiality against harvest-now-decrypt-later attacks \cite{chen2021integrated}. Further practical use cases include secure key management and distribution in large-scale networks, where QKD supplies fresh symmetric keys to classical encryption systems, as well as protection of critical control channels in applications such as smart infrastructure and industrial systems \cite{peev2009secoqc}. These examples highlight that QKD is evolving from a purely physical-layer technology to an enabler of application-driven, quantum-secure services.

The standards of the International Telecommunication Union (ITU) define multiple layers of QKD security: the physical layer handles key generation through QKD devices, a key management system (KMS) layer distributes keys between nodes, and an application layer consumes those keys for arbitrary purposes~\cite{recommendation20193800, recommendation3802, recommendation3803}. In addition to providing interfaces between QKD devices and applications, the KMS can perform functions such as key storage, synchronization, policy enforcement, and, in trusted-node networks, key forwarding between communicating parties.

Realizing fiber-based polarization QKD at scale nevertheless requires overcoming several fundamental and practical challenges. One major constraint arises from detection limitations, including finite detector efficiency, dead time, and jitter, which collectively restrict achievable key rates. In addition, optical fiber transmission introduces (i) attenuation (i.e., photon loss),  (ii) chromatic dispersion, and (iii) polarization-related effects such as polarization mode dispersion (PMD) and polarization drifts. These effects impose distinct limitations on system performance.

Attenuation affects the maximum key rate, irrespective of all else, via the Pirandola-Laurenza-Ottaviani-Banchi (PLOB) bound \cite{Pirandola2017}. Chromatic dispersion leads to a temporal broadening of wave packets, thereby degrading coincidence detection and reducing the signal-to-noise ratio over long distances~\cite{Neumann_DCM}. Polarization-related effects further complicate system stability: polarization misalignment between sender and receiver can introduce significant fluctuations in quantum bit error rate (QBER), necessitating active compensation schemes. Moreover, PMD, originating from fiber birefringence, leads to differential group delays between orthogonal polarization modes and reduces temporal indistinguishability~\cite{pmd_aps}.

While aerial fiber deployment can reduce installation and maintenance costs, it is often used primarily where aerial infrastructure already exists; in many regions, such as Germany, buried fiber tends to be preferred. Where aerial fiber is deployed, it shows considerably greater sensitivity to environmental disturbances. Wind, temperature variations and solar irradiation induce time-dependent birefringence fluctuations, leading to stronger polarization drift and increased PMD compared to buried deployments.

To extend QKD-secured communication beyond the loss-limited range, intermediate nodes, referred to as trusted nodes \cite{trusted_node}, are deployed along the fiber path. Each trusted node contains at least two QKD devices to communicate with the preceding and subsequent nodes, whether end nodes or further trusted nodes. Data are then transported hop by hop through AES tunnels keyed by the QKD-derived keys. Unlike PQC-based key exchange, this key establishment relies on physical assumptions rather than computational hardness, so the keys remain secure against an adversary with unbounded computational power. The tunnel as a whole, however, is not information-theoretically secure: because the AES key is far shorter than the payload it protects, confidentiality of the transported data remains computational. Beyond that, the intermediate nodes themselves become an attack vector, as they hold the QKD keys in plaintext. An attacker with access to a trusted node may exfiltrate secret keys or intercept communication.

To mitigate the vulnerability of trusted nodes, recent studies have explored the integration of QKD with PQC as complementary quantum-safe technologies~\cite{geitz2023hybrid, garms2024experimental, fedorov2023hybrid}. More recently, hybrid QKD--PQC architectures have been proposed for healthcare applications, where QKD-derived keys are managed through centralized KMS to secure medical data workflows~\cite{papadopoulos2026hybrid}. Both technologies can be integrated directly or combined in a layered architecture. Since PQC is not limited by transmission distance, it can provide end-to-end authentication and encryption between communication endpoints, while QKD continuously supplies fresh symmetric key material for individual network links. In conventional architectures, PQC can additionally harden the KMS responsible for relaying QKD-generated keys between network nodes.

In this work, we adopt the layered quantum-secure networking approach previously introduced by Spooren \textit{et al.}~\cite{spooren2026pqc}. In contrast to conventional QKD deployments, which commonly rely on dedicated KMS software to forward QKD-generated keys to applications, the proposed architecture directly uses QKD keys to secure hop-by-hop VPN tunnels while using end-to-end PQC between communication endpoints. The resulting layered architecture avoids dedicated KMS-based key forwarding by integrating QKD directly into a standard Linux networking stack. By leveraging established networking technologies, including conventional routing, VPNs, and firewalling, our approach simplifies deployment, monitoring, and maintenance while allowing network administrators to operate quantum-safe communication systems without specialized KMS software or expertise. Building on this framework, we demonstrate a field-deployed quantum-secure telemedicine network operating over approximately 140 km of installed telecom fiber infrastructure in Thuringia, Germany. The network enabled quantum-hardened telemedicine communication between a rural health kiosk in Sundhausen (SND) and the University Hospital Jena (\textit{Universitätsklinikum Jena}, UKJ), interconnected through a trusted-node architecture comprising Erfurt (ERF) and Jena (Fraunhofer IOF) over heterogeneous underground and aerial optical fiber links. We implemented industrial-grade entanglement-based QKD using polarization-entangled photon pairs distributed over deployed telecom fibers following the BBM92 protocol. To ensure reliable long-distance operation under realistic environmental conditions, we incorporated active polarization stabilization and chromatic dispersion compensation. We integrated the generated quantum keys into the layered security architecture, combining link-level QKD-secured tunnels with end-to-end PQC to secure telemedicine communication without requiring modifications to the underlying medical systems. 

Beyond the application demonstration, this work provides a long-term operational study of deployed entanglement-based QKD infrastructure, including 22 days of autonomous operation over a predominantly aerial fiber link. The two links, one predominantly underground and the other predominantly aerial, exhibited markedly different stability characteristics, although they were not measured concurrently. Environmental analysis further identified wind as the parameter most strongly correlated with QBER fluctuations on the aerial link. Together, these results demonstrate a practical architecture that avoids dedicated KMS-based key forwarding while integrating QKD and PQC into healthcare communication systems, and provide engineering insights into the long-term operation and robustness of real-world quantum-secure networks.

\section{System overview}

\begin{figure*}[t]
    \centering
    \includegraphics[width=0.8\textwidth]{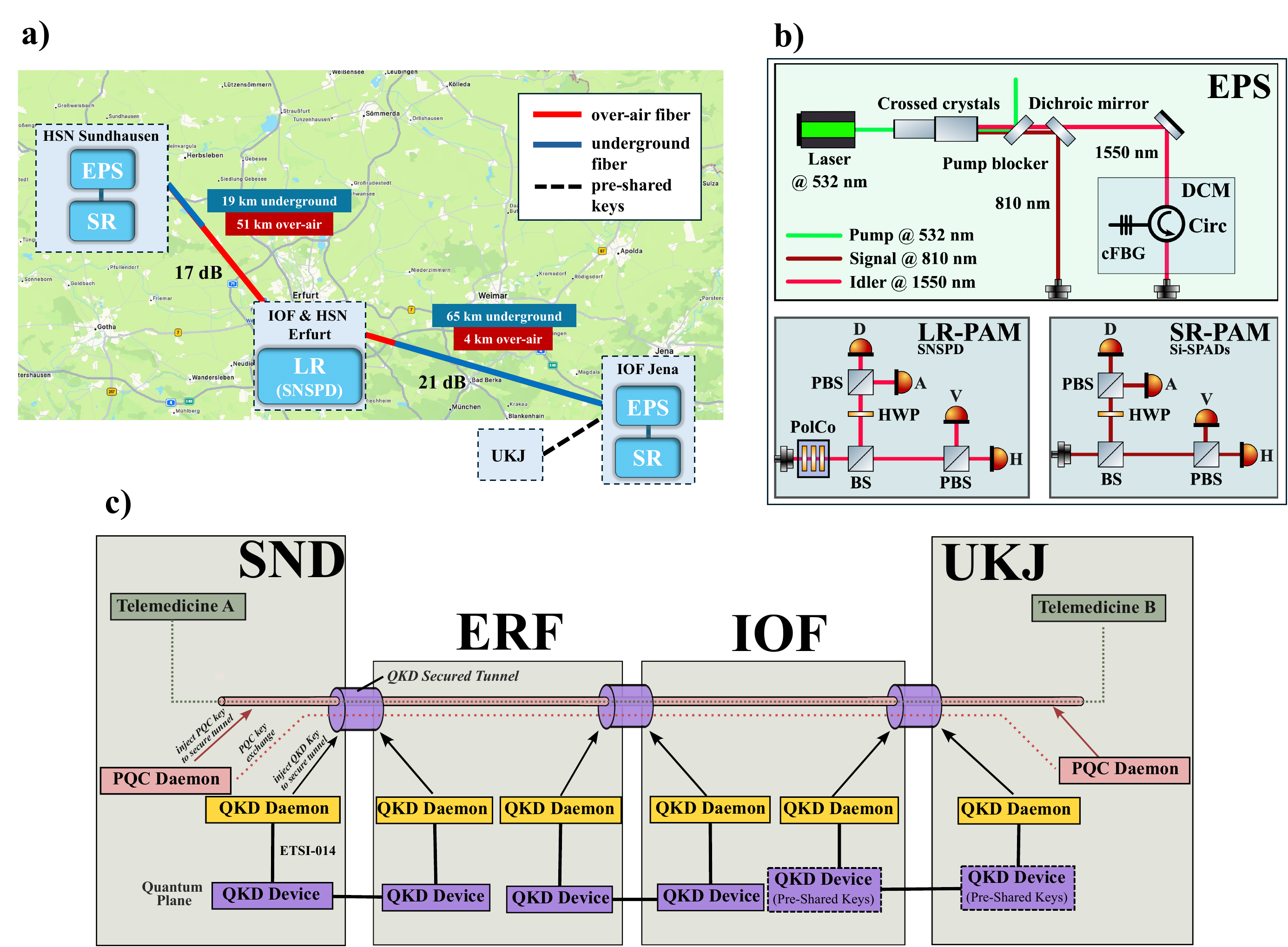}
    \caption{(a) Field-deployed entanglement-based QKD network connecting multiple nodes via underground and over-air fiber links. (b) QKD system schematic. The entangled photon source (EPS) generates photon pairs at 810 nm and 1550 nm via spontaneous parametric down-conversion (SPDC) pumped at 532 nm. The idler arm includes a dispersion compensation module (DCM) comprising a circulator (Circ) and chirped fiber Bragg grating (cFBG). Polarization analysis is performed in the short-range polarization analysis module (SR-PAM) and long-range polarization analysis module (LR-PAM) using beam splitters (BS), polarizing beam splitters (PBS), half-wave plates (HWP). The polarization drifts are compensated by the polarization controller (PolCo). Detection is realized with silicon single-photon avalanche diodes (Si-SPAD) at SR-PAM, while at LR-PAM we implemented superconducting nanowire single-photon detectors (SNSPD). (c) Layered security architecture for the telemedicine use case. The end nodes SND and UKJ host telemedicine applications and PQC daemons, while the trusted nodes ERF (Erfurt) and IOF (Fraunhofer IOF, Jena) relay QKD keys. Adjacent nodes are connected by QKD-secured tunnels (inner layer), and an end-to-end PQC-secured tunnel (outer layer) provides additional post-quantum protection between the terminal sites.}
    \label{fig:network_QKD_system}
\end{figure*}

FIG.~\ref{fig:network_QKD_system}(a) shows the real-world entanglement-based QKD network deployed in Thuringia, Germany. It interconnects the village of Sundhausen with the cities of Erfurt and Jena. In Sundhausen, we installed the system in a local \textit{Gesundheitskiosk} (i.e., health kiosk), where medical data can be securely transmitted to a clinical center in Jena. Erfurt serves as a trusted node for onward transmission to Jena. The network combines aerial and buried fiber links, reflecting realistic metropolitan and inter-city infrastructure.

The general architecture of the QKD system is illustrated in FIG.~\ref{fig:network_QKD_system}(b). We generate polarization-entangled photon pairs via type-0 spontaneous parametric down-conversion (SPDC) in a crossed-crystal entangled photon source (EPS) configuration \cite{kwiat_crossed_crystals, maroon}. The source comprises two orthogonally oriented, periodically poled potassium titanyl phosphate (ppKTP) crystals. It is pumped by a diode-pumped solid-state continuous-wave laser at 532 nm, producing highly nondegenerate photon pairs with central wavelengths of 810 nm (signal) and 1550 nm (idler).

To mitigate chromatic dispersion in the long-range arm, we implemented a dispersion compensation module (DCM). It consists of an optical circulator and a chirped fiber Bragg grating (cFBG). The DCM is slope-matched to 75 km of standard G.652 fiber ($-1260\ \mathrm{ps/nm}$ at 1550 nm). By compensating for dispersion, it improves temporal resolution, enabling shorter coincidence windows, reducing accidental coincidences, and improving the achievable secure key rate \cite{Neumann_DCM, 404km}. Although it introduces additional insertion loss, its benefit becomes dominant at longer transmission distances ($>$ 60 km), where dispersion would otherwise limit system performance.

Polarization analysis modules (PAMs) measure single-photon correlations in two mutually unbiased bases: horizontal/vertical (H/V) and diagonal/anti-diagonal (D/A). The short-range PAM (SR-PAM), located near the source, uses silicon single-photon avalanche diodes (Si-SPADs), whereas the long-range PAM (LR-PAM), positioned after the fiber link, uses superconducting nanowire single-photon detectors (SNSPD) for detection at 1550 nm. The raw key generation is based on coincidence measurements between detection events in the two PAMs. 

To compensate for polarization drifts in the fiber channels, we implemented active polarization control (PolCo) in the LR-PAM. The controller derives feedback directly from the coincidence correlations measured in the H/V and D/A bases. An algorithm adjusts the retardances to minimize the QBER and thereby realign the measurement bases of Alice and Bob. Since the entangled photons themselves serve as the reference, no auxiliary light is required, and slow polarization drifts can be tracked during key generation \cite{SansaPerna2025Polarization}.

We distill a symmetric secret key following the BBM92 QKD protocol \cite{bennett1992quantum}. After basis reconciliation, timestamps are sifted to identify coincidences within the selected time window and basis, while events with ambiguous photon-pair assignments are discarded. Errors are detected using a block-based procedure capable of identifying up to three bit errors per block; mismatched blocks are removed, and the remaining streams are verified using cyclic redundancy checks (CRC), with unsuccessful chunks likewise discarded. Finally, the disclosed information is used to evaluate the QBER, and privacy amplification, accounting for both the QBER and information leaked during processing, is applied to distil the secret key using an adapted Devetak–Winter approach.

From a networking perspective, we implemented both end nodes in SND and UKJ using embedded router hardware. Specifically, a Teltonika RUT976 and a GL.iNET Flint 2 were deployed, performing both the PQC and QKD tunnel termination and providing local network access over WiFi for connected telemedicine equipment. The nodes Erfurt (ERF) and Fraunhofer IOF Jena (IOF) were built on general-purpose x86 hardware based on AMD Threadripper processors with 4 GB of random-access memory (RAM).

All nodes run OpenWrt, a Linux-based operating system, providing a unified software environment across heterogeneous hardware platforms. QKD-secured tunnels between adjacent nodes are established as WireGuard~\cite{donenfeld2017wireguard} VPNs and hardened with QKD-derived keys. These keys are injected into the VPN by Arnika, a dedicated daemon~\cite{githubGitHubArnikaprojectarnika} that bridges the ETSI GS QKD 014 key-delivery API with the VPN software. At the PQC layer we used QuantShake\footnote{\url{https://github.com/aparcar/quantshake}} to perform the post-quantum key-exchange handshakes.

The individual network segments are interconnected by x86-based router hardware, comprising GoWin 25G and Supermicro 1U server platforms, which perform IP routing between the segmented networks. Layer-2 connectivity between network segments is provided by Mellanox Spectrum SN3700 network switches. All router hardware runs restrictive firewall configurations to limit traffic to authorized communication paths.

\section{Quantum-hardened telemedicine}

In Germany, the ongoing centralization of hospitals, combined with a growing shortage of physicians in rural areas, has reduced medical care in villages~\cite{vandenberg2019hospital, sturm2023intersectoral}. To improve healthcare accessibility without increasing travel times for medical personnel, telemedicine is gaining importance. Telemedicine uses audio and video calls to enable qualified rural staff to consult with hospitals and laboratories remotely. In addition to real-time communication, transmitted data may include medical records such as laboratory results or X-ray images. Such data are highly sensitive and require strong security guarantees. Although telemedicine systems are already in use, the vendors providing this software may not yet incorporate post-quantum security measures.

In this work, telemedicine communication between a village health kiosk (SND) and a city hospital (UKJ) was secured end-to-end using the layered approach described below; FIG.~\ref{fig:network_QKD_system}(c) shows the resulting network architecture. The two end nodes, SND and UKJ, each run a telemedicine application alongside a PQC software daemon, while the two intermediate nodes, ERF and IOF, act as trusted relays along the fiber path. Each pair of adjacent nodes is connected by a dedicated QKD-secured tunnel that encrypts the link-level traffic, and the QKD devices continuously supply fresh key material. On the IOF--UKJ segment, previously generated keys from a local keystore were used instead, owing to hardware-availability constraints. On top of these hop-by-hop QKD tunnels, the PQC daemons at SND and UKJ negotiate a single end-to-end PQC-secured tunnel; the intermediate nodes merely forward its encapsulated traffic, so this outer layer provides post-quantum authentication and encryption that is independent of the trusted nodes.

We designed the system  as a retrofit solution for existing telemedicine devices, adding the combined QKD and PQC hardening layer without requiring modifications to the downstream equipment. Furthermore, the approach is agnostic to the type of data transported, enabling seamless extension from audio and video calls to arbitrary data required by telemedicine hardware.

\section{Results}

\subsection{QKD metrics}

\begin{figure*}[t]
    \centering
    \includegraphics[width=\textwidth]{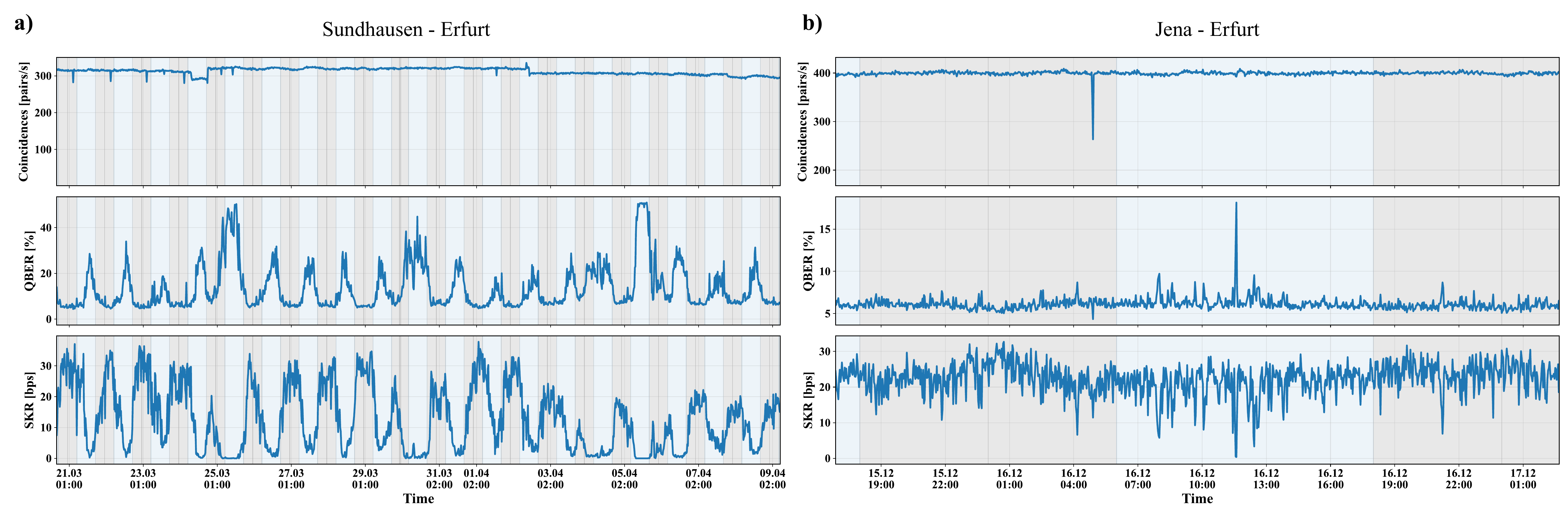}
    \caption{Long-term stability measurements of the entanglement-based QKD links between (a) Sundhausen–Erfurt and (b) Jena–Erfurt. The plots show the time evolution of coincidence rate, quantum bit error rate (QBER), and secret key rate (SKR).}
    \label{fig:metrics}
\end{figure*}

We first commissioned the IOF--ERF testbed in September 2022 using our QKD equipment. During this initial deployment, more than 300,000 quantum keys were generated over the approximately 75-km route during a ten-day measurement campaign, demonstrating sustained operation under field conditions~\cite{FraunhoferIOF2022}. In the present work, we extended this infrastructure toward Sundhausen by establishing an additional QKD link between SND and ERF, with ERF acting as a trusted node between the two links. QKD was successfully demonstrated on both links, with the corresponding performance metrics summarized in Table~\ref{tab:qkd_metrics}. Because the available QKD equipment and experimental resources did not permit simultaneous operation, the IOF--ERF and SND--ERF links were operated separately for two and 22 days, respectively. Throughout these campaigns, the active polarization-control system compensated for environmentally induced drifts without manual intervention [FIG.~\ref{fig:metrics}(a) and (b)]. The generated secret keys were stored in dedicated key pools for subsequent integration into the telemedicine proof-of-concept application.
\begin{table}[ht]
\centering
\caption{Measured QKD performance metrics for the two fiber links.}
\label{tab:qkd_metrics}
\renewcommand{\arraystretch}{1.2}
\begin{tabular}{lcc}
\toprule
\textbf{} & \textbf{Sundhausen -- Erfurt} & \textbf{Jena -- Erfurt} \\
\textbf{} & 70 km, $> 17$ dB & 69 km, $> 21$ dB \\
\midrule
Coincidences [Hz] & $311.9 \pm 15.7$ & $398.6 \pm 14.3$ \\
QBER [$\%$]       & $13.3 \pm 9.6$   & $6.1 \pm 0.8$ \\
SKR [bps]         & $12.7 \pm 10.3$  & $22.2 \pm 4.7$ \\
\bottomrule
\end{tabular}
\end{table}

The coincidence rate represents the number of temporally correlated detection events recorded per second and therefore indicates the amount of raw correlated data available for key generation. The difference in coincidence rates between the two links primarily results from the different coincidence windows used during the respective measurement campaigns and should therefore not be interpreted solely in terms of fiber attenuation.

The QBER denotes the fraction of sifted bits for which Alice and Bob obtained inconsistent outcomes when measuring in compatible bases. Such errors may arise from imperfect state preparation and polarization alignment, accidental coincidences, detector noise, and environmentally induced changes in the fiber birefringence. The QBER values reported in Table~\ref{tab:qkd_metrics} are averages over the respective measurement campaigns, while the corresponding $\pm$ values represent the temporal standard deviations. The comparatively large standard deviation for the SND--ERF link therefore reflects pronounced variations in its performance. Although its average QBER was above the threshold for secret-key generation, secret keys were generated during time intervals in which the instantaneous QBER remained sufficiently low; intervals exceeding the security threshold produced no secret key. Accordingly, the reported average SKR was calculated from the secret-key rates obtained for the individual time intervals and cannot be inferred directly from the average QBER. The lower and more strongly fluctuating SKR of the SND--ERF link is consistent with the greater environmental sensitivity of its predominantly aerial fiber route, as examined in more detail below.

\subsection{Link stability analysis}

The SND--ERF QKD link consists of 70~km of optical fiber, of which 51~km are deployed aerially, resulting in a total channel attenuation of approximately 17~dB. Due to both the relatively high losses and the over-air deployment, the link exhibits increased instability and enhanced sensitivity to environmental perturbations. To investigate the origin of these fluctuations, we analyzed the influence of several weather parameters, namely temperature, wind speed, and direct solar radiation, and correlated them with the measured QBER. Meteorological data were obtained from the Open-Meteo Service \cite{openmeteo} for a single model grid point at 50.98° N, 10.96° E, retrieved at 15 minutes resolution, with the QBER averaged onto the same grid. The wind quantity used in this section is the 10 m gust peak.

\begin{figure*}[t!]
    \centering
    \includegraphics[width=\textwidth]{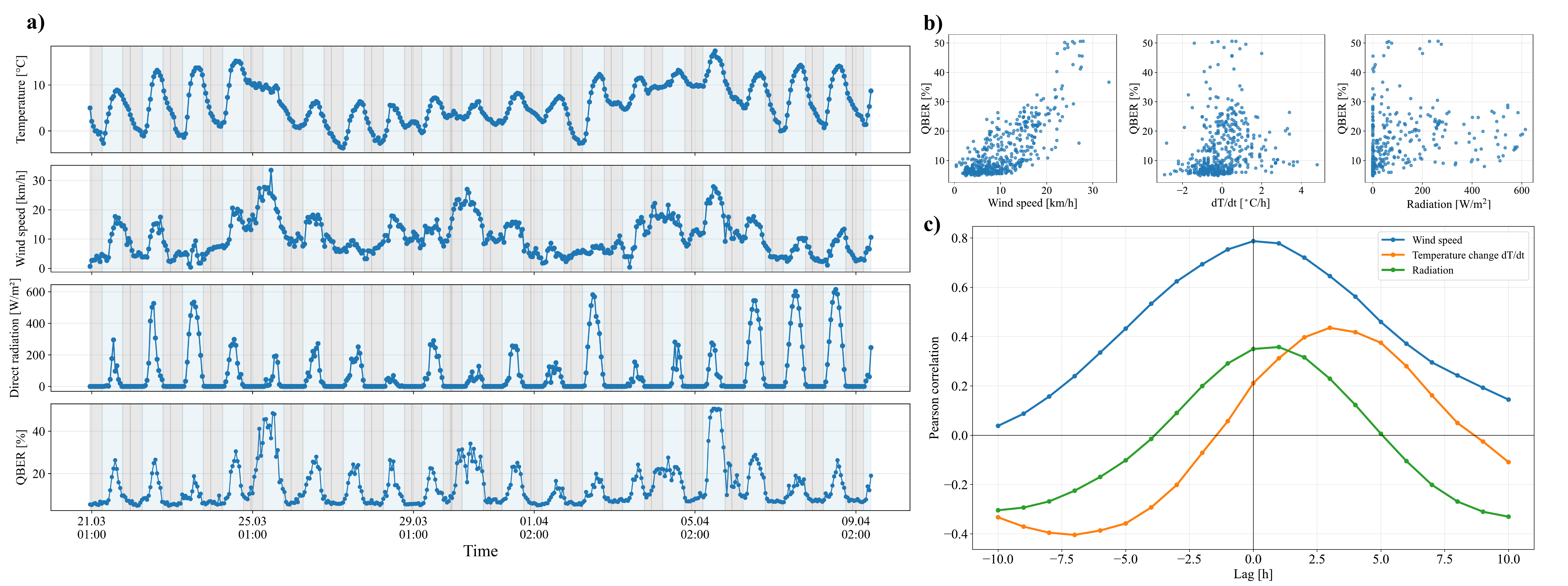}
    \caption{(a) Temporal evolution of temperature, wind speed, direct solar radiation, and QBER measured during the QKD experiment over the fiber link between Sundhausen and Erfurt. Shaded regions indicate day/night intervals. (b) Correlation of QBER with wind speed, temperature variation rate (dT/dt), and direct solar radiation. (c) Cross-correlation analysis between the QBER and meteorological parameters.}
    \label{fig:results_meteo_QBER}
\end{figure*}

As shown in Fig.~\ref{fig:results_meteo_QBER}(a) and (b), periods of stronger wind gusts generally coincided with higher QBER on the aerial-fiber link. Wind was the strongest correlate among the environmental parameters examined, with a Pearson correlation coefficient of $r =$ 0.78 at 15-min resolution ($n=$1898 intervals containing both key-generation and weather data). Averaging over longer intervals reduced short-term fluctuations and increased the coefficient monotonically to $r =$ 0.84 at six hours. We nevertheless report the 15-min result because it represents the finest temporal resolution supported by both datasets.

Because the wind-speed and QBER time series each exhibit strong temporal autocorrelation, a nominal p-value would overstate the evidence; we use a moving-block bootstrap with blocks of 12 h, above the 8.3 h decorrelation time that a Bartlett estimate over a 48 h horizon gives, and obtain a 95 $\%$ interval of [0.69, 0.84], narrowing only to [0.72, 0.83] at blocks of 96 h, long enough to hold a whole weather situation.

Wind, temperature, and solar radiation each exhibit a diurnal cycle, meaning that their raw correlations with QBER could merely reflect a shared time-of-day dependence. We therefore fitted sine and cosine components with periods of 24 and 12~h to each time series using least squares and correlated the resulting residuals. The wind correlation coefficient remained essentially unchanged, decreasing only from 0.78 to 0.75, whereas the coefficients for $dT/dt$ and solar radiation decreased from 0.26 and 0.34, respectively, to $-0.15$.

Two further checks agree: in a multiple regression on standardised predictors,  gust peak, $dT/dt$, solar radiation and the two harmonics, the wind carries the largest standardised regression coefficient, $\beta= 0.72$, at a coefficient of determination of $R^2 = 0.71$, and restricted to the night hours (22–04), when direct irradiance is zero throughout, the correlation is still 0.78.

The data suggest a transition near a gust speed of 21 km/h: the correlation was 0.33 below and 0.79 above this value. Because the breakpoint was estimated from the same dataset, this threshold should be regarded as exploratory rather than as an independently validated physical limit.

These findings are insensitive to the choice of wind metric or meteorological data source. Using the 10~m mean wind speed yields a correlation coefficient of 0.775, compared with 0.782 for the original wind metric. Moreover, 10-minute measurements from the DWD station at Erfurt--Weimar, located 4.3~km from the fiber route, reproduce the same ordering, with coefficients of 0.77 for peak gust speed and 0.71 for mean wind speed. By contrast, gustiness relative to the prevailing wind level does not show a meaningful correlation with QBER. For the gust factor, defined as the ratio of peak gust speed to mean wind speed and evaluated only above 5~km/h, where this ratio is meaningful, the correlation coefficient is $r=0.06$ for the modeled data and $r=0.22$ for the measured data, decreasing to $r=-0.10$ after controlling for the time-of-day dependence.

FIG.~\ref{fig:results_meteo_QBER}(c) presents the lag-dependent Pearson correlation between the QBER and the investigated meteorological parameters. Wind speed exhibits by far the strongest correlation, at the value quoted above. Its maximum occurs at zero lag, indicating an almost instantaneous response of the QKD link to wind-induced perturbations. In contrast, the rate of temperature change, $\mathrm{d}T/\mathrm{d}t$, exhibits a moderate correlation ($r \approx 0.43$), with a maximum at a lag of approximately $3\,\mathrm{h}$. Solar radiation shows an even weaker maximum correlation ($r \approx 0.36$), occurring after about $1\,\mathrm{h}$. These delayed responses are consistent with the thermal inertia of the fiber cable and its surrounding infrastructure, which may delay the effect of environmental heating on the fiber. However, because the correlations with dT/dt and solar radiation vanish after removing the diurnal components, the observed lags may instead reflect a shared time-of-day dependence and should not be regarded as evidence of a delayed thermal response.

The IOF--ERF link consisted of approximately \(65~\mathrm{km}\) of optical fiber deployed underground and an additional \(4~\mathrm{km}\) overhead fiber section, resulting in a total channel attenuation of approximately \(21~\mathrm{dB}\). In general, underground fiber infrastructure provides improved environmental isolation and therefore exhibits greater stability compared to aerial fiber segments. 

To compare the two links on equal footing in duration, the 39 h of stable operation of the buried link can be set against every 39 h sliding window of the aerial link: none of the 436 windows reaches the QBER standard deviation of the buried link over that window, and the quietest aerial window exceeds it by a factor of 5.1.

Monitoring of the QKD- and PQC-secured tunnel demonstrated continuous key renewal across all layers. The live QKD system generated keys at a rate exceeding the consumption of the QKD-secured tunnel, enabling uninterrupted operation. The PQC handshake software successfully renegotiated keys every 120~seconds. During the experiment, we used two different PQC algorithms (sntrup761, ML-KEM-768) to demonstrate cryptographic agility. QuantShake operated successfully over the QKD-secured network infrastructure.

\section{Discussion}

The central contribution of this work is the demonstration of application-oriented quantum-secure communication over heterogeneous real-world telecom infrastructure through the integration of entanglement-based QKD and PQC. The presented system combines long-distance field-deployed entanglement distribution, layered QKD and PQC security, trusted-node operation, and application-level integration within a realistic healthcare communication scenario. The network operated over approximately \(140~\mathrm{km}\) of mixed underground and aerial telecom fiber infrastructure and enabled secure telemedicine communication between rural and urban healthcare sites. Beyond the communication demonstration itself, the work additionally provides operational insight into the behavior of deployed quantum networks under realistic environmental conditions. These results demonstrate not only the feasibility of integrating QKD into existing communication infrastructures, but also the importance of infrastructure-dependent effects for future large-scale quantum network deployments.

The stability measurements further provide insight into the operational limitations of deployed entanglement-based QKD networks over heterogeneous fiber infrastructure. While the underground-deployed IOF--ERF link exhibited comparatively stable performance throughout the measurement campaign, the predominantly aerial SND--ERF link showed significantly enhanced sensitivity to environmental perturbations, manifested through pronounced QBER fluctuations. The lag-dependent Pearson correlation analysis provides further insight into the origin of these fluctuations. Wind speed exhibits by far the strongest correlation with the QBER, with its maximum occurring at zero lag, as reported above. In contrast, the rate of temperature change and solar radiation exhibit weaker correlations, with maximum correlation coefficients of approximately \(0.43\) and \(0.36\), occurring after delays of approximately \(3~\mathrm{h}\) and \(1~\mathrm{h}\), respectively. Although these delayed maxima are consistent with the thermal inertia of the fiber cable and its supporting infrastructure, the corresponding correlations vanish after removing the diurnal components. The observed lags may therefore arise from a shared time-of-day dependence rather than from a causal thermal response.

The observed behavior is consistent with time-dependent birefringence fluctuations and polarization mode dispersion (PMD) induced by mechanical stress and thermal expansion of the aerial fiber. Wind-driven cable motion can introduce bending, twisting, and local mechanical stress, resulting in rapid changes in the fiber birefringence and, consequently, in the polarization transformation experienced by the transmitted photons. Thermal effects, on the other hand, modify the birefringence more gradually as the cable temperature follows changes in the surrounding environment. In particular, stress points along aerial fiber sections are expected to be especially susceptible to these environmental perturbations, leading to dynamic variations in the PMD coefficient and consequently to temporal distinguishability of polarization modes. Follow-up studies combining longitudinally resolved polarimetric measurements, distributed strain and temperature sensing, and end-to-end measurements of the Jones matrix and differential group delay could localize the most sensitive fiber sections and distinguish polarization rotations from PMD-induced decoherence.

Within the present experiment, the entangled photon source operated with an emission bandwidth of approximately \(250~\mathrm{GHz}\), placing the system in a regime where PMD-induced decoherence is expected to become significant. Assuming a PMD coefficient of \(D_{\mathrm{PMD}} = 0.1~\mathrm{ps}/\sqrt{\mathrm{km}}\), as specified by the fiber provider, the corresponding polarization decoherence length is approximately \(80~\mathrm{km}\)~\cite{PMD_decoherence_2026}. Beyond this characteristic length scale, the differential group delay between the principal states of polarization becomes comparable to the photon coherence time, introducing temporal distinguishability between horizontally and vertically polarized photon pairs. This reduces the polarization entanglement visibility, increases the QBER, and ultimately limits the effectiveness of the active polarization-control algorithm. Since the deployed aerial link already approaches this regime, environmental perturbations such as wind-induced mechanical stress may contribute not only to rapid polarization fluctuations but also to variations in the effective PMD, both of which can degrade the stability of the QKD link. The observed wind-correlated QBER fluctuations are therefore consistent with a combination of the finite tracking speed of the active polarization-control algorithm and PMD-induced decoherence.

As a future step, we therefore aim to redesign the entangled photon source toward a narrower spectral bandwidth of approximately \(100~\mathrm{GHz}\). The resulting increase in photon coherence time is expected to extend the polarization decoherence length from approximately 80 km to approximately 500 km, thereby improving the robustness of entanglement distribution over environmentally exposed aerial fiber infrastructure. Moreover, we plan to implement faster polarization compensation to further improve the stability of the aerial fiber link.

The telemedicine use case was demonstrated through a 15-minute simulated consultation between a physician at a city hospital and a participant at a village health kiosk. The consultation proceeded without interruptions or perceptible delay. According to the telemedicine vendor, the multi-layered security architecture introduced no observable degradation of the audio or video quality. More generally, the presented retrofit architecture is readily transferable to other applications. Any system communicating exclusively through QKD- and PQC-secured tunnels while relying on standard TCP or UDP transport protocols can be integrated without requiring modifications to the application layer.

An important architectural advantage of the proposed approach is that it avoids dedicated KMS-based key forwarding. Instead of distributing QKD-generated keys from a centralized KMS to end applications, the generated keys are consumed directly by the VPN layer to secure each network hop, while end-to-end confidentiality and authentication are provided independently by PQC. This design reduces architectural complexity, leverages mature Linux networking technologies, and minimizes reliance on specialized QKD middleware while remaining compatible with existing operational practices.

Despite these encouraging results, several limitations remain. Due to hardware availability, the present demonstration did not rely exclusively on live QKD-generated keys but also used previously generated key material. Future experiments should therefore evaluate a fully live deployment operating solely with continuously generated QKD keys. A more detailed investigation of environmentally induced PMD is also required, particularly for aerial fiber links. Such a study should combine direct PMD characterization with simultaneous monitoring of QBER, polarization evolution, and environmental conditions to clarify the underlying mechanisms and distinguish correlation from causation. Furthermore, the demonstrated topology comprised only one village and one city end node connected through a single trusted node. Future demonstrations should consider more complex topologies involving multiple healthcare sites, enabling the evaluation of scalability, resource management, and dynamic, secure routing in practical quantum-secured communication networks.

\section{Conclusion}

We have demonstrated a field-deployed quantum-secure telemedicine network integrating entanglement-based QKD with PQC over 140 km of installed telecom fiber infrastructure in Thuringia, Germany. The implemented architecture combined trusted-node QKD links with end-to-end PQC-secured communication, enabling application-level quantum-safe telemedicine transmission between rural and urban healthcare sites without modification of the underlying medical systems.

A central novelty of this work lies in the combination of several key elements within a single operational platform: long-distance entanglement-based QKD using highly nondegenerate photons, real-world aerial and underground fiber deployment, active polarization stabilization, dispersion-compensated broadband entanglement distribution, and layered QKD plus PQC security integrated directly into a practical healthcare use case. In particular, the long-term field operation over a predominantly aerial fiber link provided important insight into the stability limitations of environmentally exposed quantum communication channels. The strong correlations observed between wind speed and QBER identify wind-induced polarization perturbations as a major practical constraint for the entanglement distribution over aerial telecom infrastructure. The measurements are consistent with contributions from rapid polarization evolution beyond the tracking bandwidth of the polarization controller and potentially from time-dependent PMD; distinguishing these mechanisms will require simultaneous polarization and PMD characterization. The buried link was substantially more stable throughout its measurement window, although fiber type, season, and equipment condition remain confounding factors.

Beyond the specific telemedicine demonstration, the presented architecture establishes a framework for quantum-secure communication in critical infrastructures. By combining information-theoretic security from QKD with the flexibility and end-to-end protection offered by PQC, the demonstrated layered approach provides resilience against both future quantum attacks and trusted-node vulnerabilities. The compatibility with standard networking hardware and conventional routing infrastructure further highlights the practical deployability of the system. In addition to the demonstrated communication performance, the proposed layered security architecture provides a practical systems-level approach for deploying quantum-safe networks. By avoiding dedicated KMS-based key forwarding and instead integrating QKD-generated keys directly into standard VPN technologies, the proposed architecture simplifies deployment, operation, and maintenance while remaining fully compatible with existing applications and communication infrastructures.

Beyond demonstrating the feasibility of field deployment, this work highlights the engineering challenges that will shape the next generation of quantum communication systems. As quantum-secure networks move toward widespread deployment, long-term robustness, environmental resilience, autonomous operation, and scalable manufacturing become increasingly important alongside security itself. The methodologies presented here for understanding and mitigating infrastructure-dependent impairments provide practical design guidelines for future industrial quantum communication systems capable of reliable operation in large-scale critical infrastructures.

\section*{Acknowledgment}

The authors acknowledge financial support from the projects Q-fiber (BMFTR, PN 16KISQ121K), AKRIT (EU/EFRE/REACT-EU, PN 2024 VFE 0089), and Q-net-Q (EU/BMFTR, PN 101091732). The presented results were partially obtained using facilities and equipment funded by the Free State of Thuringia within the Application Center for Quantum Engineering.

\section*{Author contribution}

V.-L.D.: writing - original draft preparation (lead); writing – review and editing (equal); formal analysis (lead); visualization (lead); investigation (equal); conceptualization (equal).
P.S.: writing – review and editing (equal); visualization (equal); investigation (equal); conceptualization (equal); validation (equal); software (lead).
S.M.: formal analysis (equal); validation (equal); writing – review and editing (equal).
A.Z.: supervision (equal); project administration (equal); investigation (equal); conceptualization (equal); validation (equal); writing – review and editing (equal). 
A.L.: investigation (equal); validation (equal); writing – review and editing (equal).
P.V.: project administration (equal); software (equal); writing – review and editing (equal).
M.B.: resources (equal).
J.J.: resources (equal).
S.M.: investigation (equal); validation (equal); writing-review and editing (equal).
J.D.: resources (equal).
N.P.T.: supervision (equal); project administration (equal); investigation (equal); validation (equal); writing-review and editing (equal).
F.S.: supervision (equal); funding acquisition (equal); resources (equal), writing – review and editing (equal).
K.F.: funding acquisition (equal); resources (equal); supervision (equal).
T.H.: funding acquisition (equal); resources (equal); supervision (equal), writing – review and editing (equal).
O.d.V.: funding acquisition (equal); supervision (lead); resources (equal).

\bibliographystyle{apsrev4-2}
\bibliography{references}

\end{document}